\documentclass[11pt]{article}
\pdfoutput=1

\usepackage{a4wide}
\usepackage{Packages}

\usepackage{Definitions}

\usepackage{comment}
\usepackage{setspace}

\newcommand{\OfficialTitle}{On the Renormalizability of Ho\v rava--Lifshitz--type Gravities}

\hypersetup{pdfauthor={Domenico Orlando, Susanne
    Reffert},pdftitle={\OfficialTitle}}

\author{
  \begin{minipage}{.97\linewidth}
    \vspace{1cm}
    \begin{center}
      \begin{small}
        \textbf{Domenico Orlando} and \textbf{Susanne Reffert}
      \end{small}
    \end{center}
    \vspace{1cm}
    \begin{center}
      \begin{minipage}{.7\linewidth}
        {\it \begin{footnotesize}
            \begin{center}
              Institute for the Mathematics and Physics of
              the Universe, \\The University of Tokyo, Kashiwa-no-Ha
              5-1-5, \\ Kashiwa-shi, 277-8568 Chiba, Japan.
            \end{center}
          \end{footnotesize}}
      \end{minipage}
    \end{center}
    \vspace{1cm}
  \end{minipage}
}

\date{}

\title{\vspace{1.5cm}
  \begin{huge}
    \textbf{\OfficialTitle}
  \end{huge}
}

\newcommand{\diff}{\ensuremath{\int \di t \, \di^3 x\, N  \,}}

\begin{document}

\numberwithin{equation}{section}

\begin{titlepage}
  \maketitle
  \thispagestyle{empty}

  \vspace{-13cm}
  \begin{flushright}
   IPMU-09-0053
  \end{flushright}

  \vspace{14cm}

  \begin{center}
    \textsc{Abstract}\\
  \end{center}
  In this note, we discuss the renormalizability of Ho\v
  rava--Lifshitz--type gravity theories. Using the fact that Ho\v
  rava--Lifshitz gravity is very closely related to the stochastic
  quantization of topologically massive gravity, we show that the
  renormalizability of HL gravity only depends on the
  renormalizability of topologically massive gravity.  This is a
  consequence of the \textsc{brst} and time--reversal symmetries
  pertinent to theories satisfying the detailed balance condition.
 \end{titlepage}

\onehalfspace


\section{Introduction}\label{sec:intro}

In~\cite{Horava:2009uw}, Ho\v rava studies a non--Lorentz invariant
theory of gravity in $3+1$ dimensions inspired by the \emph{Lifshitz
  model}, which has appeared in the condensed matter context (see
e.g.~\cite{Ardonne:2003p1613}). One of the main features of this
theory is that unlike Einstein gravity, it is renormalizable by
power--counting arguments. Ho\v rava--Lifshitz (HL) gravity has
sparked a lot of interest. A number of follow--up works have been
published concerning its solutions (see e.g.~\cite{Lu:2009em,
  Colgain:2009fe}) and cosmological implications (see
e.g.~\cite{Horava:2009if, Calcagni:2009ar, Kiritsis:2009sh,
  Mukohyama:2009gg, Brandenberger:2009yt,Nastase:2009nk,Cai:2009pe,Takahashi:2009wc}), but many fundamental
questions have not yet been answered.  In this present note, we
investigate the \emph{renormalizability} of this model in more detail.

Lifshitz--type models exhibit anisotropic scaling between space and
time. The amount of this anisotropy is captured by the \emph{dynamical
  critical exponent} $z$. In Lorentz--invariant theories, $z=1$, while
in the gravity theory proposed in~\cite{Horava:2009uw}, $z$ is set
equal to 3 to make the theory renormalizable by power counting. \emph{Stochastic
  quantization}~\cite{Parisi:1980ys,Damgaard:1987p1516,Namiki} proves
to be a useful tool in the study of Lifshitz--type
models. In~\cite{Dijkgraaf:2009gr}, we have argued that the (scalar)
quantum Lifshitz model can be understood as resulting from the
stochastic quantization of the free boson in one dimension less. We
have further argued that adopting the manifestly supersymmetric
formalism for stochastic quantization~\cite{PARISI:1982p1560}, which
calls for the inclusion of fermionic terms into the action, allows the
consistent study of the quantum Lifshitz model and generalizations
thereof.

In this note, we follow the same philosophy.  As already pointed out
in~\cite{Horava:2008jf}, also HL gravity can be understood as the
result of a stochastic quantization, namely of \emph{topologically
  massive gravity}~\cite{Deser:1981wh, Deser:1990bj}. Using arguments
similar to those presented in~\cite{ZinnJustin:1986eq}, we will show
that (super) Ho\v rava--Lifshitz gravity is indeed renormalizable, at least if
\emph{detailed balance} is respected, and provided that its precursor
theory, topologically massive gravity, is renormalizable, as discussed
in~\cite{Deser:1990bj}. Our main tools are the structure of the action
which is implied by the detailed balance condition, and the general
properties exhibited by theories resulting from stochastic
quantization (SQ). The main object in SQ is the \emph{Langevin
  equation}, a stochastic differential equation which governs the time
evolution of the field which is quantized.  In the process of
stochastic quantization, a $D$--dimensional Euclidean field theory is
turned into a $\left( D+1\right)$--dimensional quantum field theory,
in which a new time direction has been added. This new time direction
is necessarily on a different footing than the $D$ dimensions of the
original theory, so such a theory is in general not Lorentz
invariant. A nice property of the resulting
$\left(D+1\right)$--dimensional theory is that it automatically
exhibits a supersymmetry in the new time dimension. Even though the
resulting theory is in general not Lorentz invariant, its structure is
thus very constrained and many of its properties depend largely on the
original $D$--dimensional theory.

In fact, we will argue that the renormalizability of HL gravity rests
on the renormalizability of the underlying topologically massive
gravity.  The additional structure of HL gravity can be understood in
terms of the SQ process. We will show that this structure implied by
detailed balance, and thus of the Langevin equation, is not altered by
the renormalization group flow. To show this, we make use of the
\textsc{brst} symmetry of the theory and the time reversal symmetry of
the unrenomalized action. These two symmetries together constrain the
form of the renormalized theory. To make use of the \textsc{brst}
symmetry, we adopt a formulation of SQ which calls for the
introduction of fermionic fields. It is argued in~\cite{Namiki} that
the diagrammatic contributions of the fermions to equal--time
correlators are cancelled by contributions from lines joining
auxiliary fields to the field that is being quantized. As long as one
remains on the same time--slice, it is thus possible to completely
drop all anti--commuting fields from the action, which in this case
recovers precisely Ho\v rava's action without the fermionic fields we
are working with.

As mentioned before, we make use of the detailed balance
condition. It was argued in~\cite{Nastase:2009nk} that this condition
is phenomenologically undesirable. Our result does not extend to
HL--type theories with additional terms which break detailed
balance. For such cases, a different course must be pursued to find
arguments for their renormalizability beyond power counting.

\bigskip

The plan of this note is the following. In Section~\ref{sec:prel}, we
will briefly recall the essentials of HL gravity. In
Section~\ref{sec:stochqu}, we will review some basic facts about
stochastic quantization and apply them to the case at hand. In
Section~\ref{sec:renor}, the renormalization properties of HL gravity
are discussed. In Section~\ref{sec:conc}, we will summarize our
results.

\section{Preliminaries}\label{sec:prel}

We will be very brief in reviewing the action of HL gravity and only
give the most necessary definitions without further explanation. For
the details, we refer the reader to~\cite{Horava:2009uw}.

Since HL theory has anisotropic scaling, the spacetime $\cal M$ has
the structure of a \emph{codimension one foliation} with topology
$\setR \times \Sigma$, and the theory is designed to be invariant under
the foliation--preserving diffeomorphism group
$\mathrm{Diff}_{\mathcal{F}}(\mathcal{M})$. A Riemannian metric on
such a manifold can be decomposed \`a la \textsc{adm} into the metric
$g_{ij}$ induced along the leaves of the foliation, the \emph{shift
  variable} $N_i$ and the \emph{lapse field} $N$.

The basic objects that appear in the theory are the \emph{second fundamental form}
\begin{equation}
  \label{eq:fundform}
  K_{ij}=\frac{1}{2N}(\dot g_{ij}-\nabla_iN_j-\nabla_jN_i),
\end{equation}
and the De Witt metric on the space of metrics
\begin{equation}
\label{eq:curly_deWitt}
  \mathcal{G}^{ijkl}=\frac{1}{2}\left(g^{ik}g^{jl}+g^{il}g^{jk}\right)-\lambda\,g^{ij}g^{kl}.
\end{equation}
The parameter $\lambda$ is free, the value $\lambda=1$ would be required for the full $\mathrm{Diff}(\mathcal{M})$ invariance to hold.
The full action of HL gravity is given by
\begin{multline}
  S = \diff \sqrt{g} \, \left\{ \frac{2}{\kappa^2} K_{ij} {\mathcal G}^{ijkl} K_{kl} - \frac{\kappa^2}{2} \left[ \frac{1}{w^2}C^{ij} - \frac{\mu}{2} \left(R^{ij} - \frac{1}{2}R \, g^{ij}+\Lambda_W g^{ij}\right) \right] \right. \times\\
  \times \left. {\mathcal G}_{ijkl} \left[ \frac{1}{w^2}C^{kl} - \frac{\mu}{2} \left(R^{kl} - \frac{1}{2}R \, g^{kl} + \Lambda_W g^{kl} \right) \right] \right\}.
\end{multline}
Here, $\kappa$, $\lambda$ and $w$ are dimensionless coupling
constants. The coupling constant $\mu$ has dimension 1,
$[\Lambda_W]=2$, and $C^{ij} $ is the \emph{Cotton tensor}. The lack
of Poincar\'e invariance of the theory is reflected by the fact that
the indices $i,j, \dots$ only refer to the coordinates on $\Sigma$ and the
covariant derivative $\nabla_i$ is taken with respect to the 
metric $g_{ij}$.

A convenient rewriting of the action is obtained by introducing an
auxiliary field $B^{ij}$ and observing that the quadratic term is the
variation of the action for topologically
massive gravity with respect to the metric:
\begin{equation}
  \frac{1}{w^2}C^{ij} - \frac{\mu}{2} \left(R^{ij} - \frac{1}{2}R \, g^{ij} + \Lambda_W g^{ij} \right) = \frac{1}{\sqrt{g}} \frac{\delta S_{\text{cl}}(g)}{\delta g_{ij}} \ , 
\end{equation}
where
\begin{equation}
  \label{eq:topologically-massive-action}
  S_{\text{cl}}=\frac{1}{w^2}\int\omega_3(\Gamma)+\mu\int\mathrm d^3x\sqrt g(R-2\Lambda_W),
\end{equation}
with the gravitational Chern--Simons term given by
\begin{equation}
  \omega_3(\Gamma)=\Tr \left(\Gamma\wedge \di \Gamma + \frac{2}{3} \,\Gamma \wedge \Gamma \wedge \Gamma \right) = \varepsilon^{ijk} \left( \Gamma^m_{\phantom{m}il} \partial_j \Gamma^l_{\phantom{l}km} + \frac{2}{3} \,\Gamma^n_{\phantom{n}il} \Gamma^l_{\phantom{l}jm} \Gamma^m_{\phantom{m}kn} \right) \di^3x,
\end{equation}
where $\Gamma^l_{\phantom{l}km}$ are the Christoffel symbols.
Finally, one finds that
\begin{equation}
\label{eq:Horava-det-balance}
  S  = \diff \sqrt{g} \, \left\{ B^{ij} \left(K_{ij}+\mathcal{G}_{ijkl} \frac{1}{\sqrt{g}} \frac{\delta S_{\text{cl}}}{\delta g_{kl}}\right) - B^{ij} \mathcal{G}_{ijkl} B^{kl} \right\} \, .
\end{equation}
We will refer to actions of this form as satisfying the \emph{detailed balance condition}.

\section{HL gravity and stochastic quantization}
\label{sec:stochqu}

As already hinted in~\cite{Horava:2009uw}, it is possible to
understand HL gravity as the result of stochastically quantizing
topologically massive gravity.  In the following, we will make use of
this connection to exploit the renormalization properties of
stochastically quantized theories. We will follow the notation used
in~\cite{Damgaard:1987p1516}.

In the case of a scalar theory, stochastic quantization of a Euclidean
field theory in $d$ dimensions works as follows. We supplement the field $\phi(x)$ with
an extra time dimension $t$ (which must not be confused with the
Euclidean time $x_0$). Then we demand that the time evolution of
$\phi(x,t)$ obeys a stochastic differential equation, the Langevin
equation, which allows the relaxation to equilibrium:
\begin{equation}
  \label{eq:langevin_st}
  \frac{\partial\,\phi(x,t)}{\partial t} = -\frac{\delta  S_{\text{cl}}}{\delta \phi}+\eta(x,t),
\end{equation}
with $S_{\text{cl}}$ the Euclidean action. A stochastic equation of
this type, where the flow depends on the gradient of a function of the
field is said to satisfy the \emph{detailed balance condition}.  The
correlations of $\eta$, which is a white Gaussian noise, are given by
\begin{align}\label{eq:corr_eta}
  \braket{\eta (x,t)} = 0 \, , && \braket{ \eta (x_1,t_1)
    \eta(x_2,t_2)} =  2\, \delta ( t_1 - t_2)
  \delta^d ( x_1 - x_2 ) \, .
\end{align}
Equation~(\ref{eq:langevin_st}) has to be solved given an initial
condition at $t=t_0$ leading to an $\eta$--dependent solution
$\phi_\eta(x,t)$. As a consequence, also $\phi_\eta(x,t)$ is now a
stochastic variable. Its correlation functions are defined by
\begin{equation}
  \braket{\phi_\eta(x_1,t_1)\ldots\phi_\eta(x_k,t_k)}_\eta =
  \frac{\int\mathcal{D} \eta \, \exp \left[ -\tfrac{1}{4} \int \di^d x
      \di t \, \eta^2(x,t) \right] \phi_\eta(x_1,t_1) \ldots
    \phi_\eta(x_k,t_k)} {\int\mathcal{D}
    \eta \, \exp\left[-\tfrac{1}{4}\int \di^d x\,\di t\,\eta^2(x,t)\right]}.
\end{equation}
One of the central points for stochastic quantization is  that
equilibrium is reached for $t\to \infty$, and that
\begin{equation}
  \lim_{t\to\infty} \braket{\phi_\eta(x_1,t) \ldots \phi_\eta(x_k,t)}_\eta = \braket{\phi(x_1) \ldots \phi(x_k)},
\end{equation}
\ie that the equal time correlators for $\phi_\eta$ tend to the
corresponding quantum Green's functions.

\subsection*{Stochastic quantization of gravity theories}

In order to apply this formalism to HL gravity, we need to write down a
$\mathrm{Diff}_{\mathcal{F}}(\mathcal{M})$--invariant Langevin
equation for the quantization of the classical action in
Eq.~(\ref{eq:topologically-massive-action}). In particular, even
though the field variable is $g_{ij}$, it was argued
in~\cite{Horava:2009uw} that instead of $\dot g_{ij}$, the second
fundamental form $K_{ij}$~(\ref{eq:fundform}) should appear on the
l.h.s. of the Langevin equation to maintain
$\mathrm{Diff}_{\mathcal{F}}(\mathcal{M})$ invariance.  To reduce the
amount of possibly confusing indices, we introduce some notation (see
also~\cite{Rumpf1986}).

\begin{itemize}
\item In terms of $\mathcal{G}$, the metric on 3d space is a contravariant vector,
  \begin{equation}
    g_{ij} \eqqcolon g^I \, ;
  \end{equation}
\item variations with respect to the metric $g_{ij}$ are indicated by $\partial_I$:
  \begin{equation}
    \partial_I W(g) \eqqcolon \frac{\delta W(g)}{\delta g_{ij}} \, ;  
  \end{equation}
\item the metric on the space of metrics is expressed as a covariant metric tensor,
  \begin{equation}
    G^{ijkl} = \sqrt{g} \, \mathcal{G}^{ijkl} \eqqcolon G_{IJ} \, .
  \end{equation}
  Note the presence of the $\sqrt{g}$ term which did not appear in
  Eq.~\eqref{eq:curly_deWitt}. The metric $G_{IJ}$ is the unique metric (up
  to the choice of the parameter $\lambda$) with respect to which
  coordinate transformations of $g_{ij}$ are isometries
  (see~\cite{DeWitt:1980hx}).
\end{itemize}
Indices of type $I,J$ are raised and lowered using $G_{IJ}$
and its inverse, $G^{IJ}$,
\begin{equation}
  G_{IJ}G^{JK} = \delta_I^{\phantom{I}K} \, ,
\end{equation}
which in terms of space indices should be read as
\begin{equation}
  G^{ijmn} G_{mnkl} = \frac{1}{2} \left( \delta^i_{\phantom{i}k} \delta^j_{\phantom{j}l} + \delta^i_{\phantom{i}l} \delta^j_{\phantom{j}k} \right) \, .
\end{equation}
In the following, we will need the \emph{vielbein} on the space of metrics,
\begin{equation}
  E_A^{\phantom{A}I} E_B^{\phantom{B} J} G_{IJ} = \delta_{AB}.
\end{equation}
Now we are ready to write down the Langevin equation for HL gravity:
\begin{equation}
  \label{eq:lang_HL}
  K^I = -G^{IJ} \partial_J S_{\text{cl}} + \eta^I \, ,
\end{equation}
where $\eta$ is a noise. The measure for the
noise $\eta^I$ depends on the 3d metric $g_{ij}$, which via the
Langevin equation~(\ref{eq:lang_HL}) in turn depends on $\eta^I$. This
would introduce non--linearities in the path integral that can be
avoided if, using the vielbein, we introduce a new noise $\eta^A$ that
is actually Gaussian. Its correlators are then defined independently
of $g_{ij}$, in terms of $\delta_{AB}$. More precisely,
\begin{align}
  \label{eq:noncoord}
  \eta^A = E^A_{\phantom{A}I}\eta^I \, , &&  \eta^I = \eta^A E_A^{\phantom{A}I} \, ,
\end{align}
and
\begin{equation}\\
  \braket{ \eta^A(x,t) \eta^B(x',t') } = \frac{2}{N(t)} \, \delta(x-x') \delta(t-t') \delta^{AB} \, .
\end{equation}
(Note that indices $A,B$ in the non--coordinate basis and are raised and lowered with $\delta_{AB}$).
We can now express equation~(\ref{eq:lang_HL}) in terms of  $\eta^A$ (here expressed in the coordinates of the $D$--dimensional manifold):
\begin{equation}
  K_{ij} = - \mathcal{G}_{ijkl}\frac{1}{\sqrt g}\frac{\delta S_{\text{cl}}}{\delta g_{kl}} + \eta^{\alpha\beta}E_{\alpha\beta}^{\phantom{\alpha \beta} ij } \, .
\end{equation}

Because of the Langevin equation, the metric becomes a stochastic
function. Its generating functional is given by
\begin{equation}
  \label{eq:genfun}
  Z(J) = \braket{e^{\diff J_I g^I } }_\eta = \int \mathcal{D}\eta \, \exp \left[ -\frac{1}{4} \diff \left( \eta^A \eta_A \right) + \diff J^I g_I  \right] \, .
\end{equation}
Let $M_{I}^{\phantom{I}J}$ be the variation
\begin{equation}
  M^{I}_{\phantom{I}J} (x,t) = \partial_J \left( K^I + G^{IK} \partial_K S_{\text{cl}} -\eta^I \right) = \widetilde M^{I}_{\phantom{I}J} - \eta^A \partial_I E_A^{\phantom{A}J}\, .
\end{equation}
Then the following identity holds:
\begin{equation}
  \label{eq:identity}
  1 = \int \mathcal{D}g\,\det(M)\,\delta \left( K^I + G^{IK}\partial_K S_{\text{cl}} -\eta^I \right) \, .
\end{equation}
The two factors in Eq.~(\ref{eq:identity}) can be expressed in terms
of a bosonic auxiliary field $B$ and two fermionic auxiliary fields
$\bar \psi $, $\psi$:
\begin{align}
  & \det(M) = \int \mathcal{D}\overline\psi\,\mathcal{D}\psi\,\exp \left[\frac{1}{2}\diff \overline\psi_I M^I_{\phantom{I}J} \psi^J\right] ,\\
  & \delta \left( K^I + G^{IK}\partial_K S_{\text{cl}} - \eta^I \right) = \int \mathcal{D}B\,\exp\left[ \frac{1}{2}\diff B_I(K^I + G^{IK}\partial_K S_{\text{cl}} -\eta^I)\right].
\end{align}
Plugging the identity~(\ref{eq:identity}) into the generating
functional~(\ref{eq:genfun}), one obtains:
\begin{equation}
  Z(J)= \int \mathcal{D}\eta\, \mathcal{D}g\,\mathcal{D}\overline\psi\,\mathcal{D}\psi \mathcal{D}B \, \exp \left[- S( g, \eta, \overline \psi, \psi, B) + \diff J^Ig_I \right] \, ,
\end{equation}
where
\begin{multline}
  S(g,\eta,\overline\psi,\psi, B) = \frac{1}{2}\diff \left\{ \frac{1}{2}\eta^A\eta_A+B_I(K^I + G^{IK}\partial_K S_{\text{cl}} -\eta^I) + \right. \\
   \left. -\overline\psi_I\widetilde
  M^{I}_{\phantom{I}J}\psi^J+\overline\psi_I \eta^A \partial_J
  E_A^{\phantom{A}I}\psi^J \right\} \, .
\end{multline}
Since the measure for $\eta^A$ is Gaussian, one can perform the path integral. This is equivalent to plugging the equation of motion,
\begin{equation}
  \eta_A = E_A^{\phantom{A}I} B_I - \partial_JE_A^{\phantom{A}I} \overline \psi_I  \psi^J \, ,
\end{equation}
into the action. With this,
\begin{equation}
  Z(J) = \int \mathcal{D}g\,\mathcal{D}\overline\psi\,\mathcal{D}\psi\,\mathcal{D}B \, \exp \left[ -S(g,\overline\psi,\psi, B) + \diff J^Ig_I \right] \, ,
\end{equation}
where
\begin{multline}
\label{eq:BRST-action}
S(g,\overline\psi,\psi, B) = \frac{1}{2}\diff \left[ -\frac{1}{2} \left(E_A^{\phantom{A}I} B_I - \partial_J E_A^{\phantom{A}I} \overline \psi_I \psi^J \right) \delta^{AB} \left( E_B^{\phantom{B}K} B_K - \partial_LE_B^{\phantom{B}K} \overline \psi_K \psi^L \right) + \right.\\
  +\left.B_I(K^I + G^{IK}\partial_K
    S_{\text{cl}}) -\overline\psi_I \partial_J \left( K^I + G^{IK}\partial_K
    S_{\text{cl}} \right)\psi^J \right].
\end{multline}
Rearranging the quadratic term, using $E_A^{\phantom{A}I} E_B^{\phantom{B}K} \delta^{AB} = G^{IK}$, and rescaling $B^{ij}$, the bosonic part of the action (in space coordinates) is given by
\begin{equation}
  S_B (g, B) = \frac{1}{2} \diff \sqrt{g} \, \left[ -B^{ij} \mathcal{G}_{ijkl}B^{kl} + B^{ij} \left(K_{ij}+\frac{1}{\sqrt g} \mathcal{G}_{ijkl} \frac{\delta S_{\text{cl}}}{\delta g_{kl}} \right) \right].
\end{equation}
which recovers precisely the HL action in Eq.~(\ref{eq:Horava-det-balance}). 

It is a common feature of stochastic
quantization~\cite{Damgaard:1987p1516,Namiki} that if one considers
only equal time correlators for the metric  (\ie correlators on
the same leaf of the foliation) the diagrammatic contributions of the
fermions are exactly cancelled by the contributions from lines joining
the $B$ field to the metric. In this case one can limit oneself to
the bosonic part alone.

\subsection*{BRST invariance}

The key point of our argument is that after adding the fermions, the
action $S(g,\overline\psi,\psi, B)$ is invariant under a
\emph{\textsc{brst} symmetry} which is generated by
\begin{equation}
  \label{eq:BRST-transform}
  \begin{cases}
    \delta_\varepsilon g^I=\overline\varepsilon\,\psi^I\\
    \delta_\varepsilon\psi^I=0\\
    \delta_\varepsilon\overline\psi_I=\overline\varepsilon\,B_I\\
    \delta_\varepsilon B_I=0,
  \end{cases}
\end{equation}
where $\overline\varepsilon$ is a Grassmann variable. In fact, the variation of
$S$ is given by
\begin{multline}
  \delta_\epsilon S = - \delta_\varepsilon \left( E_A^{\phantom{A}I} B_I - \partial_JE_A^{\phantom{A}I} \overline\psi_I \psi^J \right) \delta^{AB} \left(E_B^{\phantom{B}K} B_K - \partial_LE_B^{\phantom{B}K} \overline\psi_K \psi^L \right) +\\
+ B_I \delta_\varepsilon(K^I + G^{IK}\partial_K S_{\text{cl}})-\delta_\varepsilon\,\overline\psi_I\widetilde M^{I}_{\phantom{I}J}\psi^J-\overline\psi_I\delta_\varepsilon\widetilde M^{I}_{\phantom{I}J}\psi^J.
\end{multline}
We find that:
\begin{itemize}
\item the variation of the quadratic term gives
  \begin{equation}
    \delta_\varepsilon (E_A^{\phantom{A}I} B_I- \partial_J E_A^{\phantom{A}I} \overline \psi_I \psi^J) = B_I  \partial_K E_A^{\phantom{A}I} \overline \varepsilon\, \psi^K - \overline \varepsilon\, \partial_J E_A^{\phantom{A}I} B_I \psi^J - \overline\varepsilon\, \psi^K \partial_K \partial_JE_A^{\phantom{A}I}  \overline\psi_I\psi^J = 0,
  \end{equation}
  since the first two terms cancel each other and the last term
  vanishes because $\partial_K\partial_J$ is symmetric under exchange
  of the indices and $\psi^K\psi^J$ is antisymmetric.
\item The other variations vanish, since
  \begin{equation}
    B_I\,\delta_\varepsilon(K^I + G^{IK}\partial_K S_{\text{cl}})  -\delta_\varepsilon\,\overline\psi_I\widetilde M^{I}_{\phantom{I}J}\psi^J =  B_I\,\overline\varepsilon\,\psi^J\widetilde M^I_{\phantom{I}J} -\overline\varepsilon\,B_I\widetilde M^{I}_{\phantom{I}J}\psi^I = 0 \, ,
  \end{equation}
  and 
  \begin{equation}
    -\overline \psi_I \overline \varepsilon\, \psi^K \partial_K \widetilde M^{I}_{\phantom{I}J} \psi^J = -\overline\psi_I\overline\varepsilon\,\psi^K\partial_K\partial_J(K^I + G^{IL}\partial_LS_{\text{cl}})\psi^J = 0 \, ,
  \end{equation}
  where we used the fact that $\widetilde M$ is the variation of $K^I +
  G^{IK}\partial_K S_{\text{cl}}$.
\end{itemize}
Note that the \textsc{brst} symmetry could also be reinterpreted as a
supersymmetry in the time direction by introducing a superfield
$\mathscr{G}^I = g^I + \bar \theta \psi^I + \theta \bar \psi^I +
\theta \bar \theta B $. One should nevertheless be careful because of
the presence of the $\nabla_i N_j$ terms (which are analogous to
gauge--fixing terms in the stochastic quantization of gauge
theories). Moreover, since the ``classical action'' in
Eq.~(\ref{eq:topologically-massive-action}) is not bounded from below,
supersymmetry would be spontaneously broken.

\section{Renormalization properties of HL--type gravity}
\label{sec:renor}

From the form of the action in Eq.~(\ref{eq:BRST-action}) one can read off
the dimensions of the auxiliary fields. In detail,
\begin{align}
  [B^{ij}] = 3 && [\psi_{ij}] + [\bar \psi^{ij}] = 3 && [g_{ij}] = 0 \, .  
\end{align}
This means that after renormalization, the most general form of the action is (up to a rescaling of time)
\begin{multline}
  S^{(R)} (g, \bar \psi, \psi, B ) = \diff \left\{ A^{IJ}(g) B_I B_J +
  C^{IJ}_{\phantom{IJ}K}(g) B_I \bar \psi_J \psi^K +
  D^{IJ}_{\phantom{IJ}KL}(g) \bar \psi_I \bar \psi_J \psi^K \psi^L + \right.
  \\ 
  \left. + B_I {F^{(R)}}^I(g,\,\partial g) + \bar \psi_I H^I_{\phantom{I}J}(g,\,\partial g) \psi^J + I(g,\,\partial g) \right\} \, ,
\end{multline}
where $A, C, D$ are tensors of dimension zero and depend only of the
metric $g_{ij}$. ${F^{(R)}}$ and $H$ have dimension less or equal than three and are
functions of the metric and its derivatives, and $I(g,\,\partial g)$ is a function
of the metric and its derivatives of dimension less than six.

\subsection*{Constraints from BRST}

The renormalization group flow preserves the \textsc{brst} symmetry generated by the relations in Eq.~(\ref{eq:BRST-transform}). It follows that the effective action $S^{(R)}(g, \bar \psi, \psi, B) $ satisfies the Ward identity
\begin{equation}
  \psi^I \frac{\delta S^{(R)}}{\delta g^I } + B_I \frac{\delta S^{(R)}}{\delta \bar \psi_I} = 0 \, . 
\end{equation}
In detail, by counting the powers of $B$ and $\psi$ we see that:
\begin{itemize}
\item The terms coming from the variations of $A, C $ and $D$ cancel each other.
\item The terms coming from $F^{(R)}$ and $H$ cancel each other.
\item $\delta_\epsilon I(g, \, \partial g) = 0$ which implies $I(g, \, \partial g) = 0$ up to an irrelevant constant.
\end{itemize}
The renormalized action must thus take the form
\begin{multline}
  S^{(R)} (g, \bar \psi, \psi, B ) = \frac{1}{2}\diff \left\{ - \left( {E^{(R)}}_A^{\phantom{A}I}B_I  + \partial_K {E^{(R)}}_A^{\phantom{A}I} \bar \psi_I \psi^K \right) \delta^{AB} \times \right. \\
  \left. \times \left( {E^{(R)}}_B^{\phantom{B}J} B_J + \partial_L  {E^{(R)}}_B^{\phantom{B}J}\bar \psi_J \psi^L \right) + B_I {F^{(R)}}^I (g,\,\partial g) - \bar \psi_I \partial_J {F^{(R)}}^I (g,\,\partial g) \psi^J \right\} \, .  
\end{multline}
Here we recognize the same structure of the stochastically quantized
action as in Eq.~(\ref{eq:BRST-action}). We can therefore reformulate
the problem in terms of a Langevin equation:
\begin{equation} 
\label{eq:renorm-Lang-BRST}
  {F^{(R)}}^I (g,\,\partial g) = \eta^A {E^{(R)}}_A^{\phantom{A}I} \, ,
\end{equation}
where $\eta^A$ is again a white Gaussian noise.

\subsection*{Constraints from time reversal}

In order to show that the Langevin equation above satisfies a detailed
balance condition as the initial one in Eq.~(\ref{eq:lang_HL}), we have
to make use of another symmetry. Consider the generating functional in
Eq.~(\ref{eq:genfun}),
\begin{equation}
  Z (J) = \int \mathcal{D} \eta \, \exp \left[ -\frac{1}{4} \diff \left( \eta_I \eta^I \right) + \diff \left( J_I g^I \right) \right] \, .  
\end{equation}
Substituting $\eta$ in Eq.~(\ref{eq:genfun}) using the Langevin
equation in Eq.~(\ref{eq:lang_HL}), we find that the effective action
is given by
\begin{equation}
  S (\eta) = \frac{1}{4} \diff \left( \eta_I \eta^I \right) = \frac{1}{4} \diff \left( K^I +  G^{IJ} \partial_J S_{\text{cl}} \right)  \left( K_I +  \partial_I S_{\text{cl}} \right)  \, ,
\end{equation}
where the fields are thought of as functions of $\eta$. Expanding the
product one can see that the cross term only gives a boundary
contribution:
\begin{multline}
  2\diff K^I \partial_I S_{\text{cl}} = \int \di t \, \di^3 x \,
  \left( \dot g_{ij} - \nabla_i N_j - \nabla_j N_i \right)
  \frac{\delta S_{\text{cl}}}{\delta g_{ij}} = \\ = \int \di t \, \di
  x^3 \, \left\{ \frac{\di}{\di t} S_{\text{cl}}(g) + \partial_i
  \left( N_j \frac{\delta S_{\text{cl}}}{\delta g_{ij}}
  \right) \right\} \, ,
\end{multline}
where we used the fact that $S_{\text{cl}}(g)$ preserves the
diffeomorphisms of each leaf of the foliation. Similarly in the term $K^I K_I$, the cross term gives only a boundary contribution. 
We see thus that only terms with two time derivatives remain and the unrenomalized action is
\emph{invariant under time reversal}. This property must be
preserved under the RG flow.

Starting from the Langevin equation in
Eq.~(\ref{eq:renorm-Lang-BRST}), one can similarly write down the
generating functional
\begin{multline}
  Z^{(R)} (J) = \int \mathcal{D}\eta \, \exp \left[ -S^{(R)}(\eta) + \diff g^I J_I \right] = \\ = \int \mathcal{D} \eta \, \exp \left[ - \frac{1}{4} \diff {F^{(R)}}_I(g) {F^{(R)}}^I (g) + \diff g^I J_I\right] \, .
\end{multline}
Knowing that in the initial theory $F^I = K^I + G^{IJ}\partial_J S_{\text{cl}}$, up to a rescaling of the noise, ${F^{(R)}}_I(g)$ can be rewritten as
\begin{equation}
  {F^{(R)}}^I (g) = K^I + G^{IJ} \Xi_J (g) \, .
\end{equation}
It follows that the effective action reads
\begin{equation}
  S^{(R)} (\eta ) = \frac{1}{4} \diff \left( K^I K_I + \Xi^I \Xi_I + 2 K^I \Xi_I \right) \, .
\end{equation}
In order to preserve the time reversal symmetry, $K^I \Xi_I$
must be a total derivative, which implies
\begin{equation}
  \Xi_I = \partial_I S_{\text{cl}}^{(R)} (g)\, ,
\end{equation}
where $S_{\text{cl}}^{(R)}(g)$ must preserve Lorentz invariance on the leaves.
Summarizing, we find that after renormalization, the constraints of
\textsc{brst} symmetry and time reversal imply that the time dynamics is still
described by a Langevin equation of the same form as in
Eq.~(\ref{eq:lang_HL}):
\begin{equation}
  K_{ij} = - \mathcal{G}_{ijkl} \frac{1}{\sqrt{g}} \frac{\delta S_{\text{cl}}^{(R)}}{\delta g_{kl}} + \eta_{a b} {E^{(R)}}^{a b}_{\phantom{\alpha\beta} ij} \, ,
\end{equation}
or equivalently, that \emph{the detailed balance structure of the action is
preserved under the RG flow}.

In other words, the renormalization properties of the theory in
  four dimensions are completely fixed by those of the ``classical''
theory in three dimensions -- in this case topologically massive
gravity.

\section{Conclusions}\label{sec:conc}

Ho\v rava--Lifshitz gravity is a theory of gravity constructed to be
renormalizable by power counting, even though at the price of
sacrificing Lorentz invariance at short distances. Such a model is
clearly relevant for anyone interested in the questions of quantum
gravity and has thus generated a large echo. We believe that apart
from phenomenological considerations which so far have been largely of
classical nature, the fundamental questions concerning the properties
of the quantum theory need to be addressed in order to exclude issues
of consistency.

In this note, we have studied one such question, the one concerning
the renormalization properties of HL gravity beyond power counting
arguments. In fact, our results confirm its renormalizability under
certain conditions. We make use of the fact that (super) HL gravity
can be taken to be the stochastic quantization of topologically
massive gravity. Our argument relies on the renormalizability of the
latter, which even though not strictly proven, is thought to
hold~\cite{Deser:1990bj}.  

Our reasoning can be separated into two independent parts:
\begin{itemize}
\item using the \textsc{brst}--invariant formalism of stochastic quantization we prove that the quadratic structure of the action is preserved under the RG flow;
\item we observe that theories respecting the detailed balance condition are time--reversal invariant. This further constraints the structure of the action and implies the preservation of detailed balance under the RG flow.
\end{itemize}
Note that for a theory respecting detailed balance, our proof shows that the renormalization properties entirely depend on the three--dimensional underlying action. Theories in which detailed balance is broken but the quadratic form of the action is retained still receive some protection from \textsc{brst} invariance. Our construction does not apply directly to theories in which \emph{both} the quadratic structure and detailed balance are explicitly broken.

\bigskip

The properties of HL gravity and its implications are still far from
being completely understood and this field presents a large venue for
investigation. Many fundamental questions remain to be answered.

\subsection*{Acknowledgements}

We are indebted to Simeon Hellerman for enlightening
discussions and comments on the manuscript. We would furthermore like
to thank the participants of the IPMU string theory group meetings for
directing our attention to the question investigated in this note.
The research of the authors was supported by the World Premier
International Research Center Initiative (WPI Initiative), MEXT,
Japan.

\bibliography{References}

\end{document}